\documentstyle[11pt,newpasp,twoside,epsf]{article}
\def\edcomment#1{\iffalse\marginpar{\raggedright\sl#1\/}\else\relax\fi}
\reversemarginpar

\begin{document}

\title{Unified Cloudy Models of L and T Dwarfs -- Physical Basis
of the Spectral Classification in the Substellar Regime}
 \author{Takashi Tsuji}
\affil{Institute of Astronomy, School of Science, The University of Tokyo\\
2-21-1 Osawa, Mitaka, Tokyo, 181-0015 Japan}

\begin{abstract}
Based on a simple thermodynamical argument, we proposed a cloudy 
model with a warm dust cloud deep in the photosphere. 
We showed, for the first time, that a single grid of model photospheres 
($ 800 \la T_{\rm eff} \la 2600$\,K) offers a natural explanation not 
only for the division of dwarfs cooler than M into the two distinct 
types L and T but also for the changes of the spectra and colors 
along the L -- T spectral sequence. 
\end{abstract}

\section{Introduction}

The new spectral type L is characterized by the red color possibly 
due to the dust extinction while type T by the volatile molecules
such as methane and water.  Initially, we considered a fully
dusty model (model B) and a dust-segregated model (model C)
for the L and T type dwarfs, respectively. 
However, it was difficult to explain why such different cases are
realized in different types of cool dwarfs. Also, increasing observations
on cool dwarfs could not be explained even by the use of these 
different kinds of models. We then showed that all these difficulties can
be resolved by  models with a warm dust cloud deep in the photosphere
(Tsuji 2001), and we  extended this idea to a grid of unified cloudy models
(UCMs)(Tsuji 2002). In this contribution, we reexamine  some observed data  
with our UCMs being updated with the use of the new solar carbon and oxygen
abundances by Allende Prieto, Lambert, \& Asplund (2001, 2002).

\section{Unified Cloudy Models}

In the photospheres of L and T dwarfs, dust forms at its condensation 
temperature ($T_{\rm cond}$) but grows too large to be sustained in the 
photosphere at the critical temperature ($T_{\rm cr}$). For this reason, 
only small dust grains survive in the temperature range of $ T_{\rm cr} 
\la T \la T_{\rm cond} $. This means  a formation of a dust cloud whose 
temperature is fixed at rather high (note that $T_{\rm cond} \approx 
2000$\,K) independently of $T_{\rm eff}$. Since $ T \approx 
T_{\rm eff}$ at $\tau_{\rm Ross} \approx 1$, the 
dust cloud appears at the optically thin region ($\tau_{\rm Ross} < 1$)
in L dwarfs whose $ T_{\rm eff}$'s are relatively high and at the 
optically thick region ($\tau_{\rm Ross} \ga 1$) in T dwarfs whose 
$ T_{\rm eff}$'s are lower. Then the dust  will give direct observable 
effect in L dwarfs, which in fact appear to be dusty, but the dust cloud 
is situated too deep to give significant observable  effect in T
dwarfs. We then generated a grid of non-grey  model photospheres in 
radiative-convective equilibrium under the presence of the dust cloud.

\begin{figure}
\plottwo{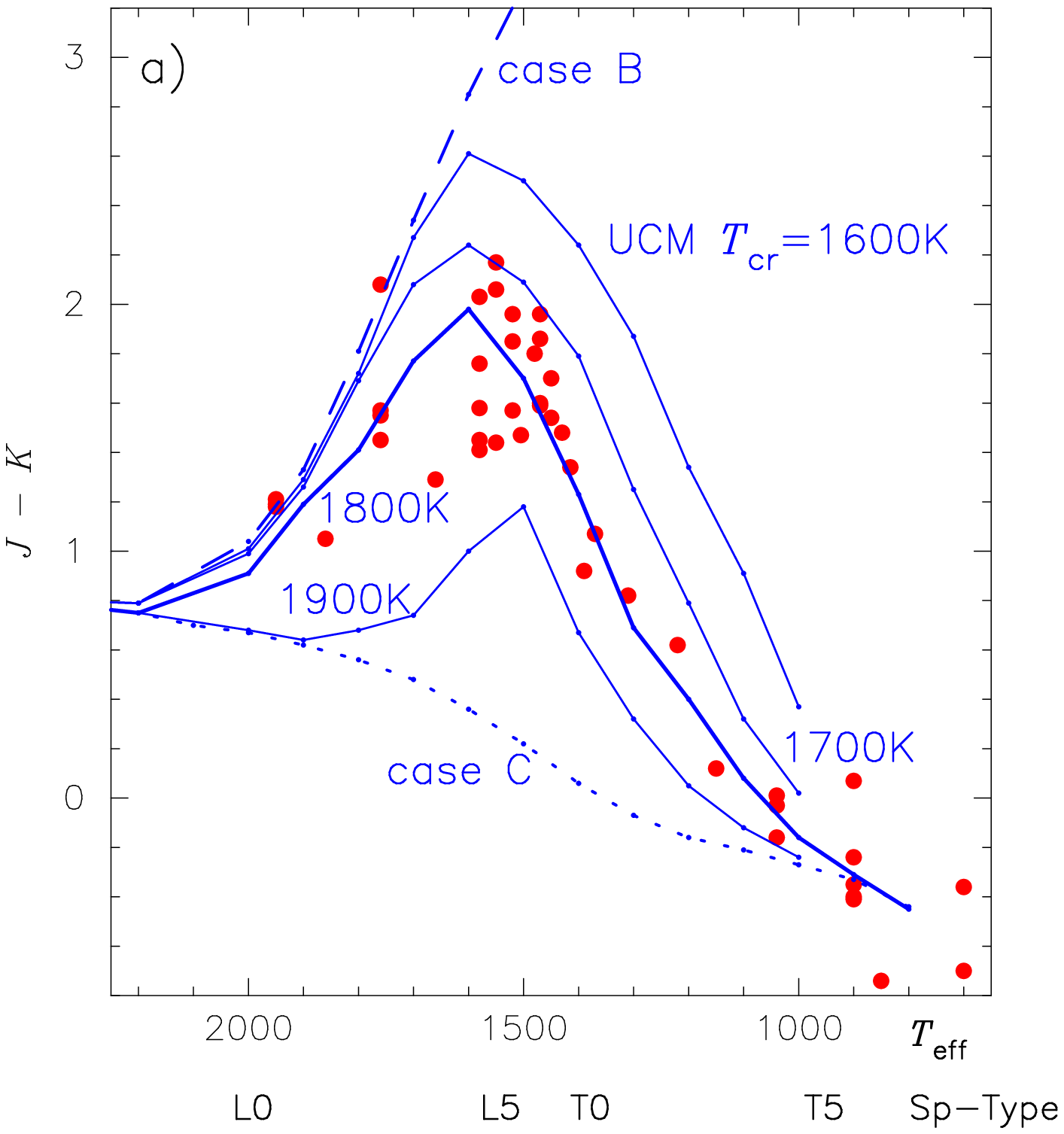}{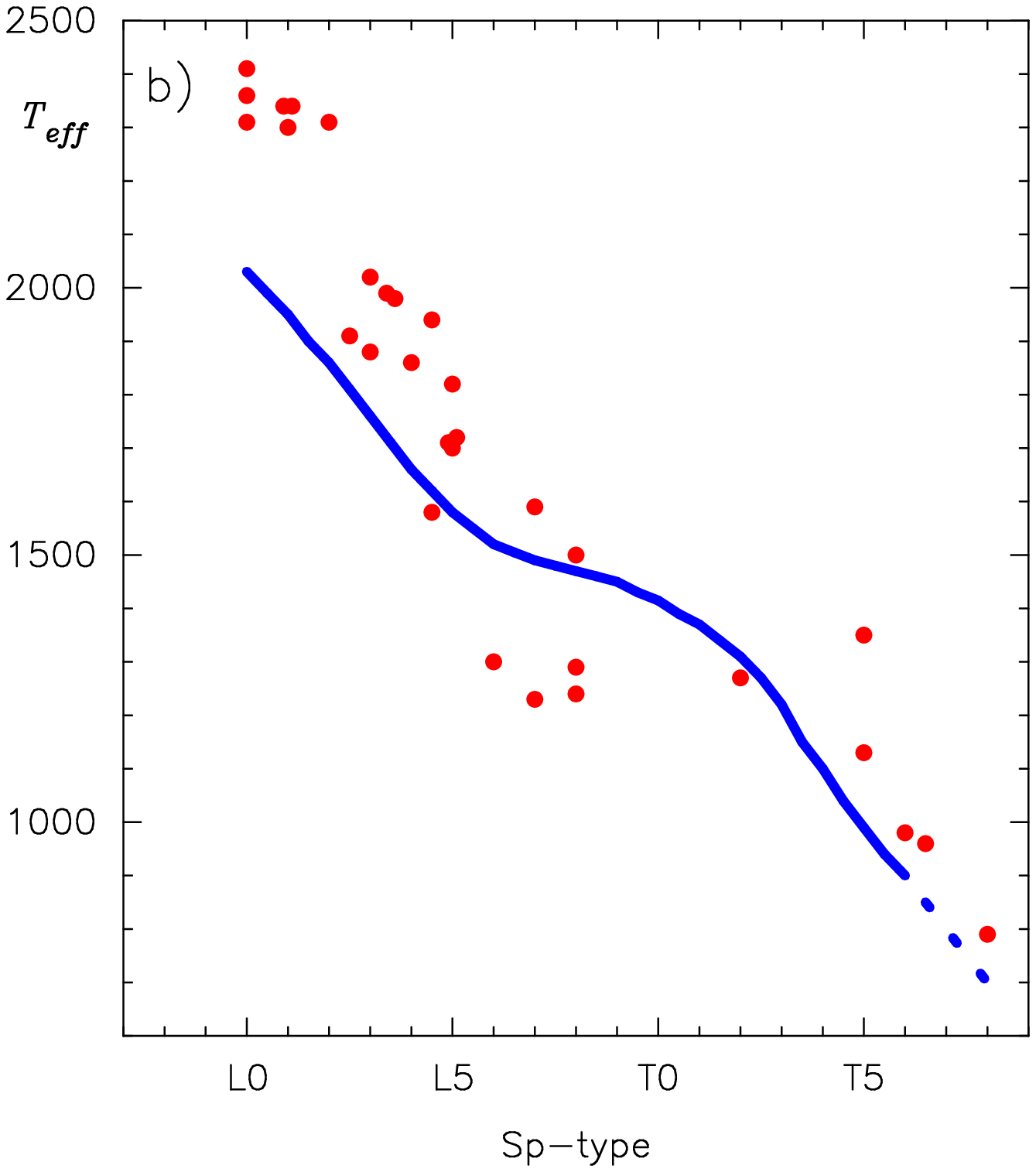}
\caption{
a) Observed $J-K$ (filled circles)  compared with the 
predicted ones based on the cloudy models with
 $T_{\rm cr} = 1600, 1700, 1800$ \& 1900\,K (solid lines),
fully dusty models B (dashed line), and dust segregated models C 
(dotted line). b) A $T_{\rm eff}$ scale based on our model analysis of 
the IR colors compared with the empirical values of $T_{\rm eff}$. 
}
\end{figure}

\section{Observational Tests of the Unified Cloudy Models }

We first analyze infrared colors based on our models. As an example,
observed values of $J-K$ (Leggett et al. 2002) plotted against spectral 
types (Geballe et al. 2002) and the predicted ones based on our 
cloudy models (UCMs) plotted against $T_{\rm eff}$ values
are compared in  Fig.\,1a. The basic feature of the observed  
color, which is redder in the later spectral types at first but turns to
blueward at  about L5, can be well accounted for by our
UCMs, but not at all by the fully dusty
models B nor by the dust segregated models C. Further,
the maximum value of $J-K$ at about L5 can be explained by our UCMs
with $T_{\rm cr} \approx 1800$\,K, and thus the critical temperature,
$ T_{\rm cr}$, can be well constrained empirically.   

Next,  effective temperatures can be estimated
so that the best fit can be obtained between the observed and predicted 
colors, and Fig.\,1a already shows the result of such a fit.
The resulting $ T_{\rm eff}$ - Sp.\,Type relationship is
shown in Fig.\,1b by the solid line (also Table 2 in Tsuji 2002).
For comparison, empirical $ T_{\rm eff}$ values  
based on the measured bolometric fluxes and parallaxes
(assuming a fixed radius) by Burgasser (2002)
are shown by the filled circles in Fig.\,1b. The agreement
is not so good especially in the early L dwarfs and further works 
should be needed before we could have a convincing $ T_{\rm eff}$ scale.   

Once the infrared colors can be  accounted for by our UCMs, 
the infrared spectra may in principle be accounted for as well.
However, some details of the molecular spectra  depend
on the chemical composition besides $T_{\rm eff}$, and
no satisfactory interpretation was possible with the use of the
solar abundances by Anders \& Grevesse (1989) 
(see Fig.12 in Tsuji 2002). Thanks to the recent
revisions not only of O but also of C abundances by
Allende Prieto et al.(2001, 2002), however,
the observed spectrum of the L dwarf prototype GD165B (Jones et al. 
1994) can be reasonably  fitted at last with the predicted one 
based on our UCM ($T_{\rm cr} = 1800$\,K) of $T_{\rm eff} =1800$\,K 
(from Fig.\,1b for this L3$\pm$1 dwarf), but never be fitted 
with those based on the models B and C, as shown in Fig.\,2.

\begin{figure}
\plotone{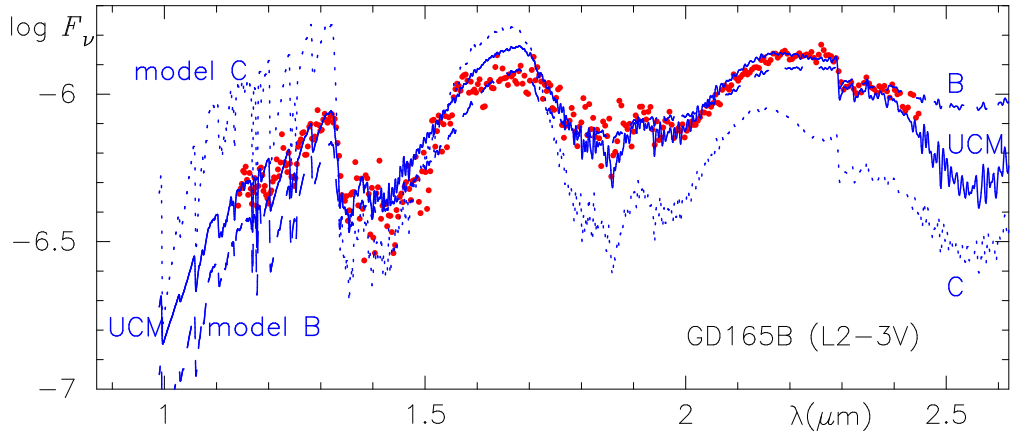}
\caption{
Observed spectrum of GD165B (filled circles) is finally fitted
with the predicted one based on the UCM  with 
log $A_{\rm C} = 8.39$ and log $A_{\rm O} = 8.69$ (log $A_{\rm H}
 = 12.0$) by Allende Prieto et al.(2001,2002)(solid line), but 
cannot at all with those based on the fully dusty model B 
(dashed line) and the dust-segregated model C (dotted line). 
}
\end{figure}

\section{Spectral Classification of L and T Dwarfs}
A unified  spectral classification of L and T dwarfs by
Geballe et al.(2002)  is based on the H$_{2}$O and CH$_{4}$
indices derived from the near IR spectra. We compare 
one of them (H$_{2}$O 1.2\,$\mu$m index) with the predicted ones
from our model spectra in Fig.\,3a. Also, 
observed EWs of K\,I 1.2432\,$\mu$m (Burgasser 2002) are compared 
with the predicted ones in Fig.\,3b.
The rather modest increase of H$_{2}$O index in the L dwarf regime
is due to the compensation of the increasing H$_{2}$O abundance
and the increasing dust extinction towards the later L dwarfs, in
which the dust cloud is still in the optically thin photosphere.
K\,I EWs show the minimum at about L5 because the dust column
density of the cloud in the optically thin region is
the largest at about L5 (or $ T_{\rm eff} \approx 1500$\,K).
After the dust cloud penetrates into the optically thick region
at $ T_{\rm eff} < 1500$\,K, the region above the cloud is
dominated by the increasing amount of volatile molecules and mildly
volatile atoms (e.g. K) towards the later T dwarfs. This explains the
rapid increase of the H$_{2}$O indices as well as the upturn of  
K\,I EWs in the T dwarf regime. 

Thus, our UCMs will provide the physical basis for the spectral 
classification of L and T dwarfs. It is to be noted, however, that 
what matters in the L -- T classification  is the formation of the dust 
cloud at different optical depths $\tau_{\rm Ross}$ (though at nearly
fixed temperature near $T_{\rm cond}$)  rather than the amount 
of dust predicted by the thermochemical theory. This shows a marked 
contrast to the classical spectral types O -- M, which are directly related 
to the amount of ions, atoms, and molecules predicted by the 
ionization and dissociation theory.

\begin{figure}
\plottwo{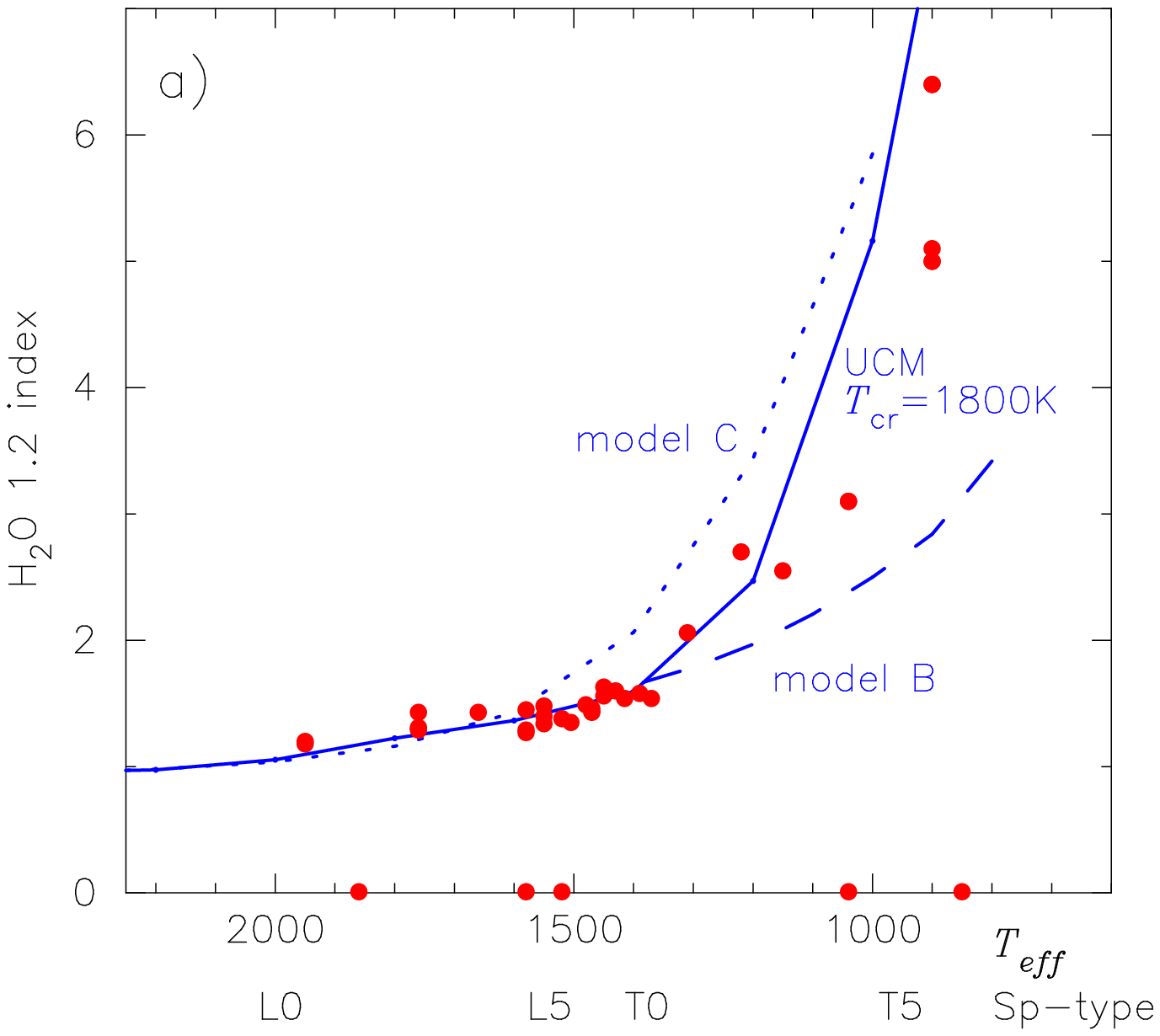}{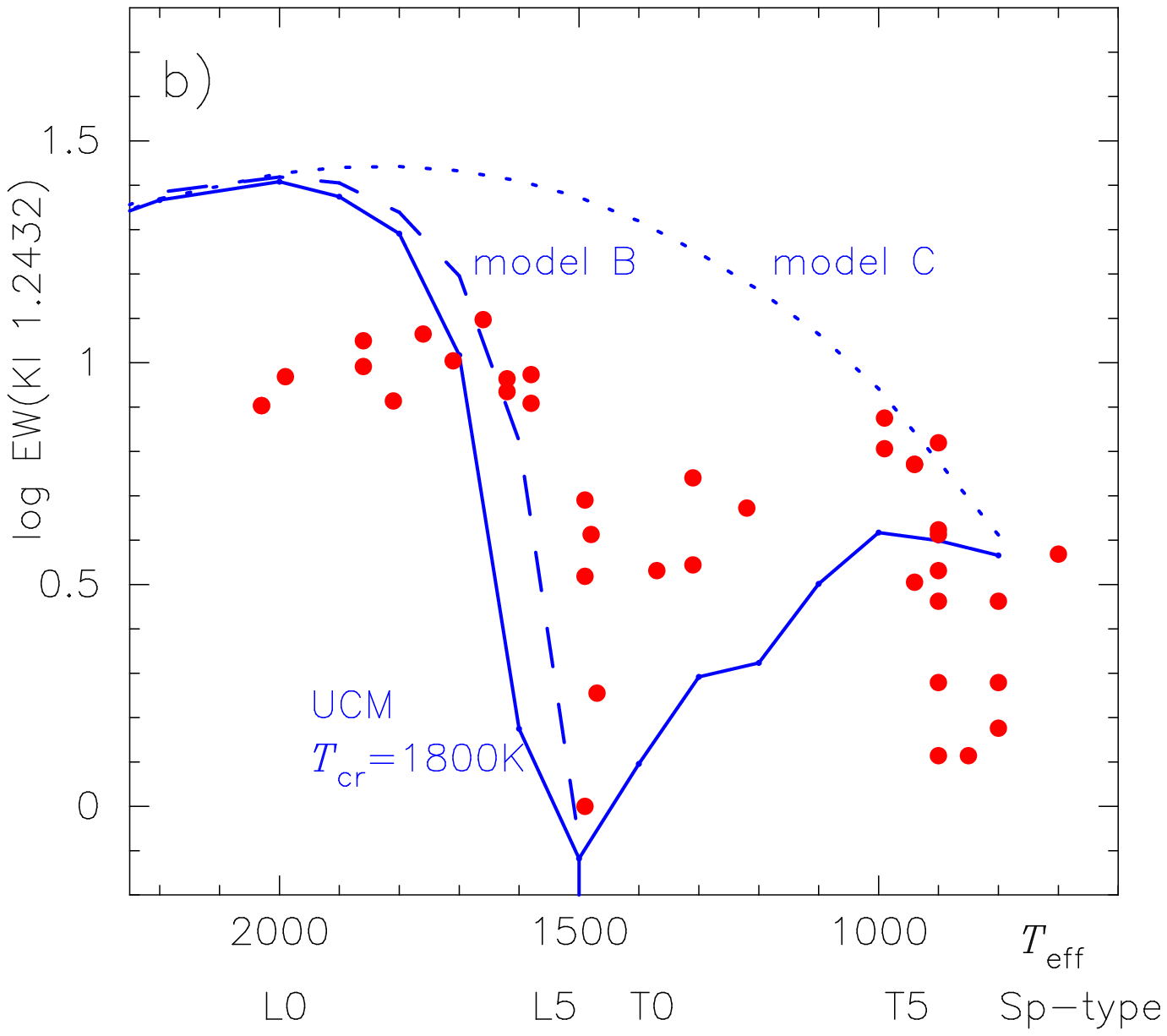}
\caption{
a) Observed H$_{2}$O index (filled circles) and 
predicted ones based on UCMs (solid line), dusty (dashed line) and 
dust-segregated (dotted line) models. b) The same as a) but 
for K\,I 1.2432\,$\mu$m EWs.
} 
\end{figure}

\section{ Concluding Remarks}
We showed that the observed spectra and colors of L and T dwarfs
can be accounted for consistently with our UCMs. 
On the other hand, our previous models such as
the fully dusty models B and the fully dust-depleted models C 
(and, by implication, more or less similar models  by other authors)
 are far too short to explain the observed colors (Fig.1a), spectra (Fig.2)
and spectral indices (Fig.3) of cool dwarfs, especially from middle L to 
early T. Thus the presence of the dust cloud deep in the photosphere
should be an essential feature of L and T dwarfs, as also
noted by Marley et al.(2002) from a different approach
based on the planetary theory. We followed stellar approach throughout
and pursued a simple model as far as possible, but it is clear that this 
is only an initial step towards understanding the complicated
phenomena in ultracool  dwarfs.

\end{document}